\documentclass[11pt,twoside]{article}
\usepackage{asp2014}

\newcommand\vir{CU\,Vir}
\newcommand\ori{V901\,Ori}

\newcommand{\zav}[1]{\left(#1\right)}
\newcommand{\hzav}[1]{\left[#1\right]}

\newcommand{\oc}{\textit{O-C}}

\aspSuppressVolSlug
\resetcounters

\begin{document}

\title{Towards comprehension of the variability of the magnetic chemically peculiar star CU~Virginis (HD 124\,224)}
\author{Z. Mikul\'{a}\v{s}ek,$^1$ J. Krti\v{c}ka,$^1$ A. Pigulski,$^2$ G. W. Henry,$^3$ and J. Jan\'{i}k$^1$}
\affil{$^1$Department of Theoretical Physics and Astrophysics, Masaryk University, Kotl\'a\v{r}sk\'a 2, CZ 611\,37 Brno, Czech Republic; \email{mikulas@physics.muni.cz}}
\affil{$^2$Astronomical Institute, University of Wroc{\l}aw , Wroc{\l}aw, Poland}
\affil{$^3$Center of Excellence in Information Systems, Tennessee
State University, Nashville, Tennessee, USA}

\paperauthor{Sample~Author1}{Author1Email@email.edu}{ORCID_Or_Blank}{Author1 Institution}{Author1 Department}{City}{State/Province}{Postal Code}{Country}
\paperauthor{Sample~Author2}{Author2Email@email.edu}{ORCID_Or_Blank}{Author2 Institution}{Author2 Department}{City}{State/Province}{Postal Code}{Country}
\paperauthor{Sample~Author3}{Author3Email@email.edu}{ORCID_Or_Blank}{Author3 Institution}{Author3 Department}{City}{State/Province}{Postal Code}{Country}

\begin{abstract}
The upper main sequence stars CU Virginis is the most enigmatic object among magnetic chemically peculiar (mCP) stars. It is an unusually fast rotator showing strictly periodic light variations in all regions of the electromagnetic spectrum, as well as spectroscopic and spectropolarimetric changes. At same time, it is also the first radio main-sequence pulsar. Exploiting information hidden in phase variations, we monitored the secular oscillation of the rotational period during the last 53 years. Applying own phenomenological approach, we analyzed 37\,975 individual photometric and spectroscopic measurements from 72 data sources and improved the O-C model. All the relevant observations indicate that the secular period variations can be well approximated by the fifth degree polynomial.
\end{abstract}

\section{Introduction}

Magnetic chemically peculiar (mCP) stars are the most suitable test beds for studying rotation and its variation in the upper (B2V to F6V) main-sequence stars. The surface chemical composition of these objects uses to be very uneven. Overabundant elements are, as a rule, concentrated into large spots persisting there for decades to centuries. The abundance unevenness of the atmospheres influences the stellar spectral energy distribution. As the star rotates, periodic variations in the spectrum, brightness, and longitudinal magnetic field are observed.  We have studied both present and archival observations of all kinds to check the stability of the rotation periods of mCP stars.

The changes rotation periods were derived from shifts of (light, phase) phase curves by means of the method developed by \citet{mik901}. They applied this method at first to the helium strong star \ori. Then, it was many times improved and tested on mCPs and other types of variables \citep[see, e.g.][]{mikecl}. The method is based on the usage of suitable phenomenological models of phase curves of rotation tracers and the period variation \citep[one can find a detailed manual in][]{mikmon}.  Solution through robust regression provides us with all model parameters and estimations of their uncertainty.

The vast majority of CP stars studied to date display strictly constant rotational periods. However, a few mCP stars, including \vir\ and \ori, have been discovered to exhibit rotational period variations caused by yet unknown reasons.

\section{Period variations of CU Virginis}

\vir\ = HD 124224, is a bright, rapidly-rotating ($\overline{P}=0.520694$\,d), medium-aged silicon mCP star \citep{kochba}. It is also the first known main-sequence stellar radio pulsar \citep[][and references therein]{trig}. \citet{pyp}, using their new and archival photometry, constructed an O-C diagram showing a sudden period increase of 2.6\,s (slower rotation) in 1984! Another smaller jump toward a longer period in 1998 was reported by \citet{pypad}.  \citet{mikCU} processed all available measurements of \vir\ and found that its rotation was gradually slowing until 2005 and since then has been accelerating.
\begin{figure}
\centering\includegraphics[width=0.63\textwidth]{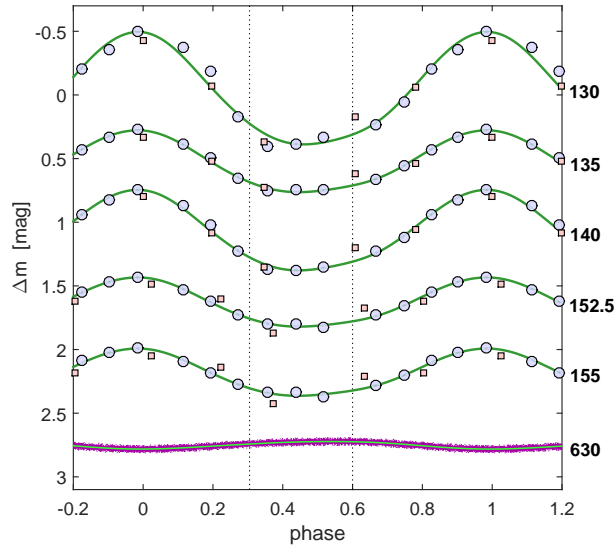}
\caption{Comparison of light curves in far UV and red regions of \vir\ spectrum. Blue circles and red squares denote magnitudes derived from IUE and HST spectra, violet points are the measurements from SMEI satellite. Wavelengths are expressed in nanometers and phases are calculated using ephemeris (\ref{OC5}). Green lines are the fits of the light curve phenomenological model, assuming two symmetric photometric spots with centers at the phases 0.303 and 0.598 (black dotted lines).}\label{LCs}
\end{figure}

\subsection{Models of phase function and period}\label{modelfun}

The first attempts to describe and model the apparent changes of the rotation period of \vir\ was based on the assumption that period changes \citep{pyp,pypad}, which can be represented as a series of linear fragments in the \oc/phase shift diagrams. The possible physics of abrupt changes was afterward discussed in \citet{step}.

\citet{mikCU} showed that the change of the period is more likely gradual, without any jumps. The phase function (sum of phase and epoch) $\vartheta(t)$ was in their paper approximated by an aperiodic, three-parametric, symmetric biquadratic function, resembling a segment of a simple cosine function.
\begin{figure}
\centering\includegraphics[width=0.65\textwidth]{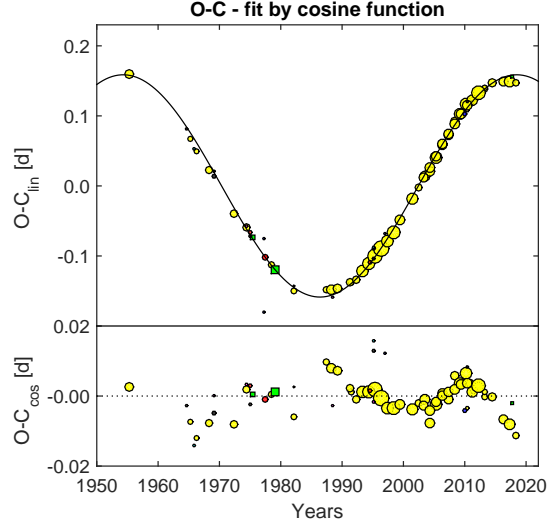}
\caption{Upper: $\oc_{\rm{lin}}$ versus time in years, where $O$ is the moment of the zero phase and $C_{\rm{lin}}=M_0+P_0\times E$, where $M_0,\,P_0$ are the parameters described in Sect.\,\ref{cucos}, and $E$ is the number of rotational cycles elapsed from the initial epoch $M_0$. Inconstancy of $\oc_{\rm{lin}}$ is the result of rotation period $P$. $\oc_{\rm{lin}}$ is fitted by a four parametric cosine function. Bottom: The residuals $\oc_{\cos}$ show apparent undulations. Optical photometry -- yellow circles, UV spectrophotometry - green squares, red circles -- equivalent widths (EWs) of \ion{He}{i} lines, black ones -- \ion{Si}{ii}, violet ones -- \ion{H}{i}, blue -- effective magnetic field, and cyan -- radial velocity measurements. The size of symbols is are proportional to adopted weights.}\label{OCcos}
\end{figure}
\citet{krt} showed that cyclic oscillations in the rotational period revealed by \citet{mikCU} might result from the interaction of the internal magnetic field and differential rotation and predict a rotational cycle timescale of 51\,yr. In that sense, we first used the model assuming periodic variations of the period $P(t)$ and phase function $\vartheta(t)$ with the period $\mathit{\Pi}$. Using all 18\,267 observations of \vir\  available up to 2015, we found the period $\mathit{\Pi}=67.6(5)$\,yr, close to the theoretical prediction.

Adopting all data known by the end of 2016 \citep[especially those of][]{pyper2}, \citet{mikmon} applied this model again in the following form:
\begin{eqnarray}
&\displaystyle \vartheta_1(t)=\frac{t-M_0}{P_0};\quad \phi(t)= \frac{t-T_0}{\mathit{\Pi}};
\quad \Delta(\phi)=-\frac{A}{P_0}\cos\zav{2\,\pi \phi};\quad \vartheta=\vartheta_1-\Delta(\phi);\label{cucos}\\
&\displaystyle \theta(\vartheta)=M_0+P_0\,\vartheta+P_0\,\Delta(\phi);\quad P(t)=P_0\frac{\mathrm{d}\vartheta_1}{\mathrm{d}\vartheta}
\doteq P_0\hzav{1+\frac{2\,\pi\,A}{\mathit{\Pi}}\,\sin\zav{2\,\pi\,\phi(t)}},\nonumber
\end{eqnarray}
where $\Delta$ is an auxiliary function, $A$ is a semiamplitude of \oc\ changes with the minimum at $T_0$ and the semiamplitude of the mean period undulation being $A_{\mathrm P}=2\,\pi AP_0/\mathit{\Pi}$. $M_0$ was chosen so that $\Delta(\tilde{\vartheta_0}=0)=\mathrm d\Delta/\mathrm d \vartheta_0=0$. Analysing all the available observational data of \vir, we found $M_0=2\,446\,604.4390$ (fixed), $P_0=0.520\,694\,04(3)$\,d. $T_0=2\,446\,604(13),\ \mathit{\Pi}=24\,110(150)\,\mathrm d=66.0\pm0.4$\,yr, $A=0.1611(5)$\,d, and $A_{\mathrm P}=1.888$\,s. The data (19\,641 individual observations) cover more than one cycle of the proposed sinusoidal variations.

At the same time \citet{mikmon} modeled the data using polynomial model of the phase function $\vartheta(t)$ and found that fourth order polynomial model (5 parameters) gave a bit better results than the harmonic one. The application of the fifth order polynomial (6 parameters) has occurred being unsubstantiated.

\section{Recent results}

\subsection{New observation}\label{recent}
Recently, the volume of the photometric data of CU Vir has been increased by the space photometry made in the years 2003-2011 by the Solar Mass Ejection Imager (SMEI) experiment \citep{eyles,jackson}. The photometry, available through the University of California San Diego web page\footnote{http://smei.ucsd.edu/new\_smei/index.html} has been processed to remove the instrumental effects. The corrections included subtraction of the repeatable seasonal variability, and subsequent rejoection of outliers and detrending. In effect, we obtained 19226 individual photometric observations for further analysis.

During 2017-8 we obtained 10 new spectrograms taken by far UV spectrograph on board Hubble Space Telescope. We yielded 25 magnitudes in 5 passbands centered at 130, 135, 140, 152.5, and 155 nm. In addition, we derived 55 new magnitudes from 11 spectrograms taken by IUE. Phased light curves of all above mentioned data are depicted in Fig.\,\ref{LCs}. 689 high-precision BV measurements acquired by one of us (GH) with the Automated Photometric Telescope (APT) at Fairborne Observatory in Arizona at 2017 and 2018 seasons harbored our \vir\ data in the present.

Presently, we have in our disposal altogether 37\,975 relevant observations of \vir\ covering sufficiently the time interval 1955-2018. The prevailing source of information is the photometry with 37\,313 measurements done/derived in the filters with the centers in the interval of 135--765 nm. The other tracers which allow to monitor period changes are measurements of equivalent widths of \ion{He}{i}, \ion{Si}{ii}, and \ion{H}{i} lines (569), effective magnetic field (59), and radial velocities (59). The present data are rich enough to improve the model of the phase function.

\subsection{Discussion of phase function models}\label{pol5}

First, we applied the simple four-parametric cosine phase function model described by relations (\ref{cucos}), used in \citet{krt} and \citet{mikmon}. The found model parameters were slightly, but significantly shifted versus those ones found before: $P_0=0.520\,693\,87(2)$\,d, $A=-0.1587(3)$\,d, $T_0=2\,446\,573(2)$, and $\mathit{\Pi}=23\,370(90)\,\mathrm{d}=64.2(3)$\,yr. The quality of the \oc\ fit can be quantified by the relative $\chi^2_r$, where we found unacceptably high value of 28. 
\begin{figure}
\centering\includegraphics[width=0.65\textwidth]{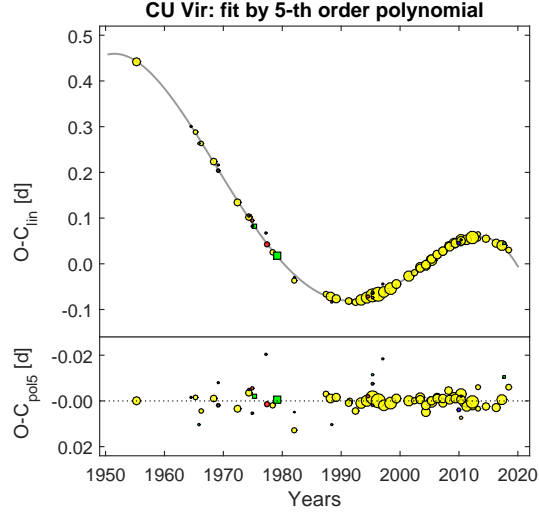}
\caption{Upper: $O-C_{\rm{lin}}$ ($C_{\rm{lin}}=M_{01}+P_1\times E$, where $M_{01},\,P_1$ are the parameters of the linear approximation, see in Sect.\,\ref{pol5}, and $E$ is the number of rotational cycles elapsed from the initial epoch $M_{01}$) versus time in years fitted by fifth-order polynomial phenomenological model. Bottom: The residuals $\oc_{\rm{pol5}}$ (see Eq.\,(\ref{pol5})) show only a weak undulation. The symbols are the same as in Fig.\,\ref{OCcos}.}\label{OC5}
\end{figure}

The detailed inspection of \oc\ diagram (Fig.\,\ref{OCcos}) shows that the cosine model can be regarded only as the first approximation of the reality. As the $\oc_{\cos}$ exhibits apparent double wave during the cycle of 65 years, we conclude that it is necessary to raise the number of free parameters describing the phase function model by at least two. It suggests itself the usage of the second-order harmonic polynomial or the fifth-order polynomial ($N=5$). Although the first possibility is permitted by theory \citep{krt}, we shall discuss the second one which is less biased.
\begin{eqnarray}
&\displaystyle\vartheta_1=\frac{t-M_{01}}{P_1},\quad \theta_1=\frac{\vartheta_1}{10^4},\quad  \displaystyle\theta=\frac{\vartheta}{10^4},\quad \frac{\mathrm{d}\theta}{\mathrm{d}\vartheta}= \frac{\mathrm{d}\theta_1}{\mathrm{d}\vartheta_1}=10^{-4},\\
&\displaystyle\vartheta(\vartheta_1)=\vartheta_1+\sum_{i=2}^N \alpha_i\,\left(\theta_1^i +\sum_{j=0}^{i-1}\beta_{ij}\theta_1^j\right),\nonumber\\
&\displaystyle T(\vartheta)\doteq M_{01}+P_1\vartheta- P_1\,\sum_{i=2}^N \alpha_i\,\left(\theta^i +\sum_{j=0}^{i-1}\beta_{ij}\theta^j\right)\nonumber\\
&\displaystyle P(t)=\frac{\mathrm{d}t}{\mathrm{d}\vartheta}= P_1\left(\frac{\mathrm{d}\vartheta}{\mathrm{d}\vartheta_1}\right)^{-1}\doteq
P_1\left[1-\frac{\mathrm{d}\theta_1}{\mathrm{d}\vartheta_1}\sum_{i=2}^N \alpha_i\left(i\,\theta_1^{i-1} + \sum_{j=1}^{i-1}j\,\beta_{ij}\,\theta_1^{j-1}\right)\right],\nonumber\\
&\displaystyle P(\vartheta)=\frac{\mathrm{d}T}{\mathrm{d}\vartheta}\doteq P_1\left[1-\frac{\mathrm{d}\theta}{\mathrm{d}\vartheta}\sum_{i=2}^N \alpha_i\left(i\,\theta^{i-1}+\sum_{j=1}^{i-1}j\,\beta_{ij}\,\theta^{j-1}\right)\right],\nonumber\\
&\displaystyle \dot{P}(t)=
\frac{1}{P_1}\frac{\mathrm{d}P}{\mathrm{d}\vartheta_1}
=-\left(\frac{\mathrm{d}\theta_1}{\mathrm{d}\vartheta_1}\right)^{\!2}\sum_{i=2}^N \alpha_i\left[(i^2\!-\!i)\,\theta_1^{i-2}+
\sum_{j=2}^{i-1}(j^2\!-\!j)\,\beta_{ij}\,\theta_1^{j-2}\right],\nonumber
\end{eqnarray}
where $M_{01}=2\,452\,652.9973(6),P_1=0.520\,703\,479(14)$\,d are ephemeris of the linear approximation $(N=1)$. For $N=5$, $\alpha_{2}=0.12069(19)$, $\alpha_{3}=-0.02914(18)$, $\alpha_{4}=-0.03039(14)$, $\alpha_{5}=-0.00223(15)$ are the dimensionless parameters founded iteratively by weighted robust regression \citep[for details see][]{mikmon}. Rightfulness of the highest order of the polynomial follows from the fact that $|\alpha_5|/\delta \alpha_5=15$. Coefficients $\beta_{20}=0.52219$, $\beta_{21}=-1.0691,\ \beta_{30}=0.82694,\ \beta_{31}=0.88898,\ \beta_{32}=-2.6527,\ \beta_{40}=0.34705,$ $\beta_{41}=3.0039,\ \beta_{42}=-0.97123,\ \beta_{43}=-3.771,\ \beta_{50}=-0.96466,\ \beta_{51}=2.1025,$ $\beta_{52}=6.2491,\ \beta_{53}=-3.362,$ and $\beta_{54}=-4.8325$ are orthogonalization parameters found by operations described in \citet{mikmon}.

\section{Conclusions}

Although $\chi^2_r=7$ of the fit of the polynomial is four times smaller than for the cosine one, its high value shows that the 5-th order polynomial fit is unable to describe observed tiny changes on the time scale of several years - see Fig.\,\ref{OC5}. Nevertheless, the last global model represents a substantial improvement with respect to the previous models and leads us to a better comprehension of variability of the rotation period of \vir.

\acknowledgements This research was supported by grant GA\,\v{C}R  16-01116S. This research was partly based on the IUE data derived from
the INES database using the SPLAT package. AP acknowledges the support from the National Science Centre grant no. 2016/21/B/ST9/01126.


\end{document}